\documentclass[11pt]{article}
\usepackage[a4paper,margin=2.5cm]{geometry}
\usepackage[T1]{fontenc}
\usepackage{lmodern}
\usepackage{amsmath,amssymb,graphicx}
\usepackage{booktabs}
\usepackage[table]{xcolor}
\usepackage{enumitem}
\usepackage[hidelinks]{hyperref}
\setlist{leftmargin=18pt}
\newcommand{\that}{\hat{t}}
\newcommand{\ctime}{c\textsubscript{time}}
\newcommand{\wmae}{\textup{WMAE}}

\title{Beyond AUC: Do Mammography Risk Models Predict When Cancer Occurs?}
\author{Svetlana Krasnova \and Andrey Krylov \and Dmitry Sorokin\\[1ex]
\normalsize Faculty of Computational Mathematics and Cybernetics,\\
\normalsize Lomonosov Moscow State University, Moscow, Russia}
\date{}

\begin{document}
\maketitle

\begin{abstract}
Longitudinal deep learning models for mammography-based breast cancer risk
prediction output either per-year cumulative risks or a discrete distribution
over the year of diagnosis, from which cumulative risks $F_1,\dots,F_5$ are
derived. These models are evaluated almost exclusively with year-wise AUCs and the concordance index.
We show that these metrics are largely insensitive to \emph{when} cancer occurs:
they can be high even for a model that carries no information about the year of diagnosis. We re-evaluate two recent longitudinal models (OA-BreaCR,
LMV-Net) on CSAW-CC with a 5-year horizon using timing metrics
computed on cancer cases only. Despite C-indices up to 0.75, neither
orders cancer cases by time to diagnosis significantly better than chance
(\ctime{} 0.51 and 0.55, both 95\% CIs covering 0.5). Weighted absolute timing
error favours a constant ``year~3'' prediction over both models, so we
use \ctime{} as the primary timing metric. For Mirai's additive-hazard head, also used by LMV-Net, we analyse
how per-year increments passed through a ReLU can collapse under severe class
imbalance. We also propose a breast-level, two-stage model
that separates \emph{whether} from \emph{when}. It matches LMV-Net in
discrimination (C-index 0.745 vs.\ 0.746) but orders cancer cases better
(\ctime{} 0.58 vs.\ 0.51, paired 95\% CI of the difference 0.01--0.13), on par
with Mirai's published weights (0.57).
\end{abstract}


\section{Introduction}
\label{sec:intro}

A breast cancer risk model can tell \emph{whether} a woman will develop cancer
and \emph{when}. Risk-adapted screening trials such as
WISDOM~\cite{esserman2025wisdom} use only \emph{whether}: they assign women
to screening schedules by 5-year risk category. Per-year outputs can add
\emph{when}, e.g., if diagnosis is likely before the next screen, but only
if curve shapes differ between women.

Mirai~\cite{yala2021mirai} introduced an
additive-hazard (``cumulative probability'') layer that outputs monotone
cumulative risks $F_1,\dots,F_5$ for the next five years. Recent longitudinal
models add prior exams~\cite{lee2023prime,karaman2024lomar} and explicit alignment. OA-BreaCR~\cite{wang2024oabreacr}
aligns current and prior feature maps with attention and uses an ordinal head
that outputs a probability for each of years 1--5 and for staying cancer-free.
LMV-Net~\cite{thrun2026lmvnet} jointly models
aligned CC and MLO views. It uses Mirai's additive-hazard layer.
Both target risk stratification, and OA-BreaCR also predicts time to
cancer~\cite{wang2024oabreacr}.

All of these works report AUC$_k$ for ``cancer within $k$ years'' and the
C-index~\cite{uno2011cstat}. Mirai~\cite{yala2021mirai}, both evaluated
methods~\cite{wang2024oabreacr,thrun2026lmvnet} and the
benchmark protocol of~\cite{thrun2025impact} count same-year diagnoses
(cancer diagnosed in the exam's calendar year) as ``cancer within one year'',
a common convention under which AUC$_1$ partly measures detection. Both metrics are dominated by
separating future cancers from controls and ignore whether the \emph{shape} of
the risk curve, the only part encoding timing, is informative. The
C-index is also known to be improper for evaluating $t$-year predicted
risks~\cite{blanche2019cindex}. Our contributions are:
\begin{enumerate}
  \item A simple invariance argument and a pair count showing that AUC$_k$ and C-index ignore
        timing and can be high for ``flat'' predictions.
  \item An analysis of how the ReLU-based additive-hazard head of Mirai and LMV-Net can collapse to flat predictions under class imbalance.
  \item A re-evaluation of OA-BreaCR and LMV-Net on CSAW-CC
        without same-year diagnoses, showing near-chance timing, and an explanation of why
        year errors do not penalise collapse.
  \item A breast-level two-stage model (cancer probability, then one timing
        score) with significantly above-chance timing that exceeds LMV-Net's at a
        comparable C-index and matches Mirai's published weights.
\end{enumerate}

\section{Methods}
\label{sec:methods}

\subsection{Problem setup}
Let $x$ be a screening exam (current and prior mammograms) and
$\tau\in\{1,\dots,5\}$ the year of diagnosis for exams followed by cancer
within 5 years (\emph{cases}); all other exams are \emph{controls}. All models
are evaluated through non-decreasing cumulative risks
$F_k(x)\approx P(\tau\le k\mid x)$. \emph{Additive-hazard heads} (LMV-Net) output $F_k$ directly (Eq.~\eqref{eq:hazard}).
\emph{Discrete-distribution heads} (OA-BreaCR) output probabilities
$(q_1,\dots,q_5,q_{>5})$ over the year of diagnosis and ``cancer-free within
5 years'', and $F_k=\sum_{j\le k}q_j$.
AUC$_k$ is the cumulative/dynamic time-dependent AUC~\cite{heagerty2000roc}: it scores $F_k$ with cases $\tau\le k$ and controls observed beyond $k$,
so later cancers count as controls. Timing is read from the curve as the
conditional distribution of the diagnosis year given an event within 5 years
(uniform if the curve has no mass), and its mean:
\begin{equation}
  p_k = F_k - F_{k-1},\quad
  g_k = \frac{p_k}{\sum_{j=1}^{5}p_j},\quad
  \that = \sum_{k=1}^{5} k\,g_k .
  \label{eq:that}
\end{equation}
For a discrete-distribution head, $g_k=q_k/(1-q_{>5})$ is the model's own
conditional distribution of the diagnosis year.

\subsection{Why AUC$_k$ and C-index are blind to flat predictions}
Predictions are \emph{flat} if $F_k(x)=a_k\,\phi(x)$ with a scalar risk
$\phi$ and constants $0<a_1\le\dots\le a_5$. In words, every woman gets the
same curve shape, scaled by her risk.
\textbf{Proposition.} \emph{For flat predictions, (i) each AUC$_k$ and the
C-index (evaluated on any $F_\tau$) equal those of $\phi$; (ii) $g$, and hence
$\that$, is identical for all $x$, so $\ctime=\tfrac12$ exactly.}
\emph{Proof.} (i) Both metrics depend on the scores only through their ranking
(the IPCW weights depend only on times), and $a_k\phi$ ranks like $\phi$.
(ii) $g_k=(a_k-a_{k-1})/a_5$ does not depend on $x$, and ties count $\tfrac12$.
$\square$ The statement involves only $F_k$, so it holds for both head types.

The C-index does contain case--case pairs with different $\tau$, but they are
rare. Counting unweighted pairs with $n_1$ cases and $n_0$ controls followed
for 5 years, at most $\binom{n_1}{2}$ of the $\binom{n_1}{2}+n_1n_0$
comparable pairs are case--case pairs. On our test set ($n_1{=}188$, $n_0{=}7{,}210$) this is ${\le}1.3\%$. Perfect timing can therefore
change the C-index by less than $0.007$ relative to random within-case
ordering.

\subsection{Collapse of ReLU-based additive-hazard heads under class imbalance}
\label{ssec:collapse}
The additive-hazard head used by
LMV-Net~\cite{thrun2026lmvnet} originates from
Mirai~\cite{yala2021mirai}. It adds a linear baseline and non-negative
per-year increments, each a linear layer followed by a ReLU:
\begin{equation}
  F_k(\mathbf f)=\sigma\Big(b(\mathbf f)+\textstyle\sum_{j\le k}
  \mathrm{ReLU}\big(\mathbf w_j^{\top}\mathbf f+c_j\big)\Big).
  \label{eq:hazard}
\end{equation}
With
2.5\% positives, most of the gradient pushes all $F_k$ down. The baseline $b$
can follow freely, but the increments are bounded below by zero, so the
pre-activations are driven negative. A ReLU unit that is negative for (almost)
all inputs receives zero gradient and rarely recovers~\cite{lu2020dyingrelu}. If all increments become
inactive, then $F_1=\dots=F_5=\sigma(b(\mathbf f))$: a flat prediction with
$\that\equiv1$. AUC$_k$ and C-index are unaffected, while timing is lost
entirely. On our test set, the increment is zero for all exams in years 2--4 for LMV-Net. From year~1 to year~5 its risk rises by only 0.21 percentage points on average (from $\approx$1.2\%), i.e.\ the curve is nearly flat. Mirai's published weights show an analogous partial collapse: the year-3 and year-5 increments are zero for all 5{,}011 test exams. This does not apply to OA-BreaCR (no per-year ReLU).

\subsection{Timing metrics}
\label{ssec:metrics}
Timing metrics use cases only. \textbf{\ctime}, our primary metric, is the fraction of correctly
ordered case pairs with different true years,
$\mathcal P=\{(i,j):\tau_i<\tau_j\}$, possibly across women:
\begin{equation}
  \text{\ctime}=\frac{1}{|\mathcal P|}\sum_{(i,j)\in\mathcal P}
  \Big(\mathbf{1}[\that_i<\that_j]+\tfrac12\,\mathbf{1}[\that_i=\that_j]\Big).
\end{equation}
Unlike the time-dependent concordance~\cite{antolini2005ctd}, it excludes
case--control pairs.
\textbf{\wmae} averages $|\that_i-k|$ within each true year $k$ and then
across years, so that frequent years do not dominate.
\textbf{WCS} does the same with $\mathbf{1}[|\that_i-k|\le1]$ (higher is
better).
\textbf{Year-error metrics do not penalise collapse.} \wmae{} and WCS,
including the variants used by OA-BreaCR~\cite{wang2024oabreacr}, reward
assigning everyone a central year: $\that\equiv3$ gives $\wmae=1.2$ and
WCS${}=0.6$ for any case distribution, and the nearly flat LMV-Net obtains a
relatively high WCS (Table~\ref{tab:main}), whereas $\ctime=0.5$ for any
constant prediction. We therefore report \wmae{} and WCS only alongside this
constant reference. Our implementation of these metrics differs from the public
OA-BreaCR evaluation code~\cite{wang2024oabreacr}, where (a)~$\that=\sum_k k\,q_k$ includes the
cancer-free bin (index 5), pulling $\that$ towards 5, whereas we condition on
cancer (Eq.~\eqref{eq:that}); (b)~controls are included in the error, with
per-class weights, whereas we use cases only, averaged over true years; (c)~errors are rounded
before thresholding (an error of 1.4 counts as a hit), whereas we do not round.

\subsection{Breast-level labels and two-stage model}
\textbf{Labels.} The original baseline pipelines assign a
woman's cancer label and $\tau$ to both breasts, so the healthy contralateral
breast inherits the other breast's time to cancer. We instead use
breast-level labels for all methods, including the baselines: only the breast
that develops cancer is a case and the contra\-lateral breast is cancer-free.

\begin{figure}[t]
\centering
\includegraphics[width=0.6\linewidth]{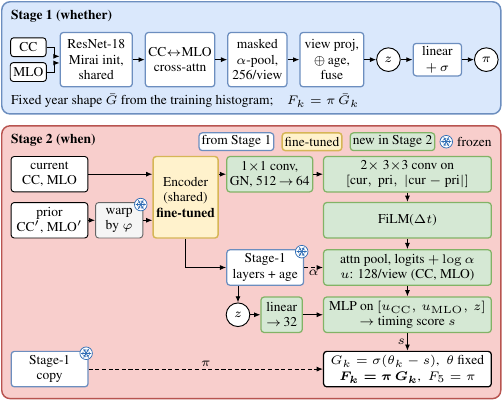}
\caption{Two-stage model (one breast). Stage~1 predicts $\pi$ with a fixed
year shape $\bar G$. Stage~2 adds priors warped by a fixed prior-to-current
deformation field $\varphi$ from our registration method~\cite{krasnova2026jiim} and learns a timing score $s$.
$\pi$ comes from a frozen Stage-1 copy. The embedding $z$ is computed by the frozen Stage-1 layers.}
\label{fig:arch}
\end{figure}

\textbf{Stage 1 (whether)} (Fig.~\ref{fig:arch}). A sigmoid-linear head
gives $\pi(x)=P(\tau\le5\mid x)$ per breast from the current CC+MLO and an age group
(${<}55$ / ${\ge}55$; no priors). The baselines do not use age. The year shape is fixed to the case-year CDF of the 316 training
cases, $\bar G=(0.27,0.60,0.75,0.92,1)$, so $F_k=\pi\bar G_k$: the image sets
only the height. The loss is the likelihood of $\pi$: $-\log\pi$ for cases
and $-\log(1-\pi\bar G_C)$ for controls followed $C$ years (class-mean
losses weighted $\tfrac14$ for cases and $\tfrac34$ for controls).
\textbf{Stage 2 (when).} A head outputs one scalar $s(x)$ (larger $s$: later
cancer) and is trained only on women who develop cancer (632 breast-exams, 316
cases, 202 women; current and prior CC and MLO). With fixed
$\theta_k=\mathrm{logit}\,\bar G_k$ (so $s{=}0$ reproduces $\bar G$) and $\pi$
from a frozen copy of Stage~1,
\begin{equation}
  G_k=\sigma(\theta_k-s),\ G_5=1,\ g_k=G_k-G_{k-1},\ F_k=\pi G_k,
  \label{eq:stage2}
\end{equation}
with $G_0=0$ and $\that=\sum_k k\,g_k$. Thus $s$ only moves \emph{when} given cancer within
5 years; \emph{whether} ($F_5=\pi$) is decided by Stage~1, and no per-year
activation can zero timing. Each step uses all exams of 16 women and minimises
\begin{equation}
\begin{split}
  \mathcal L={}&\textstyle\sum_{\tau_i<\tau_j}\mathrm{sp}(s_i-s_j)(\tau_j-\tau_i)^2
  +0.2\sum_i(\that_i-\tau_i)^2\\
  &\textstyle-\sum_c\log\big(1-\pi_c\,G_{C_c}(s_c)\big),
\end{split}
\end{equation}
where $\mathrm{sp}$ is softplus, $i,j$ are cases in the group (within a breast
or across women) and the terms are averaged. The cancer-free contralateral
breast $c$ at the same exam is a censored observation with $C_c$ years of
follow-up (capped at 5; no gradient at $C_c{=}5$), not a ranking negative.
Both stages: AdamW; training augmentation by flip ($p{=}0.5$), brightness
gain $U(0.85,1.15)$ and shift $U(-0.05,0.05)$.
Stage~1: lr $5{\cdot}10^{-5}$ (encoder) and $10^{-4}$ (head), wd $10^{-4}$,
${\le}50$ epochs, early stopping on validation C-index (patience 15). Stage~2:
head lr $3{\cdot}10^{-4}$ (wd $10^{-2}$), encoder $10^{-5}$ (wd $10^{-4}$),
${\le}80$ epochs, patience 20 on validation \ctime. The C-index depends
mainly on $\pi$, which Stage~2 keeps frozen, so Stage~2 is selected by its
target, \ctime.

\begin{table}[t]
\centering
\caption{CSAW-CC, 5-year horizon. Discrimination metrics used in prior work
(left) and timing metrics on cancer cases only (right). Concordance and AUC in
\%, errors in years. 95\% bootstrap CIs in parentheses. Bold marks the best per column among
models trained on CSAW-CC.
$^{\dagger}$Breast-level evaluation (CC+MLO fused, or CC and MLO predictions averaged for OA-BreaCR).
$^{\ddagger}$External reference: published weights, not trained on CSAW-CC.}
\label{tab:main}
\setlength{\tabcolsep}{3.0pt}
\resizebox{\linewidth}{!}{%
\begin{tabular}{l c ccccc | c c c}
\toprule
 & & \multicolumn{5}{c|}{AUC$_k$ (\%) $\uparrow$} & & & \\
Method & C-index $\uparrow$ & 1y & 2y & 3y & 4y & 5y
 & \ctime{} $\uparrow$ & \wmae{} $\downarrow$ & WCS (\%) $\uparrow$ \\
\midrule
Constant $\that\equiv3$ (ref.) & -- & -- & -- & -- & -- & --
 & 50.0 & \textbf{1.20} & \textbf{60.0} \\
\midrule
OA-BreaCR$^{\dagger}$~\cite{wang2024oabreacr} & 69.1 (64.5--73.7) & 66.1 & 71.1 & 72.4 & 71.9 & 74.6
 & 54.7 (49.6--60.1) & 1.48 (1.45--1.50) & 42.1 (40.0--45.0) \\
LMV-Net$^{\dagger}$~\cite{thrun2026lmvnet} & \textbf{74.6} (70.3--78.6) & \textbf{72.3} & 76.9 & 76.2 & \textbf{76.2} & 74.5
 & 50.9 (45.5--56.7) & 1.52 (1.36--1.70) & 45.9 (40.9--51.0) \\
\midrule
Ours$^{\dagger}$ & 74.5 (70.2--78.7) & 72.2 & \textbf{77.0} & \textbf{76.6} & 75.1 & \textbf{75.0}
 & \textbf{58.1} (51.9--64.2) & 1.39 (1.25--1.56) & 44.0 (38.9--49.7) \\
Stage~1 only$^{\dagger}$ & 73.4 (69.4--77.5) & 69.2 & 75.2 & 75.1 & 74.4 & \textbf{75.0}
 & 50.0 (50.0--50.0) & 1.31 (1.31--1.31) & 40.0 (40.0--40.0) \\
\midrule
Mirai$^{\dagger\ddagger}$~\cite{yala2021mirai} & 74.5 (70.4--78.4) & 74.7 & 76.7 & 76.1 & 75.3 & 74.9
 & 57.1 (51.4--62.2) & 1.48 (1.37--1.59) & 44.0 (39.9--47.5) \\
\bottomrule
\end{tabular}}
\end{table}

\section{Experiments}
\label{sec:experiments}

\textbf{Data.} Women in CSAW-CC~\cite{dembrower2020csaw} (Hologic, 2008--2015)
are split 50/20/30 (strictly patient-disjoint: no woman is in two splits; seeded, stratified by earliest
time to cancer). $\tau$ is the diagnosis year, inferred from the last exam's
radiological timing, minus the exam year; controls need ${\ge}5$ years of
follow-up. The prior is the preceding CC+MLO exam of the same breast (gap
mostly 2, 1--8 years).
\textbf{Exclusion of same-year diagnoses.} Breast-exams with $\tau=0$ (154/60/91 in
train/val/test) are excluded and all baselines retrained, so $\tau\in\{1,\dots,5\}$. Cases/controls (breast-exams): train
316/12{,}125, val 129/4{,}838, test 188 (from 124 women)/7{,}210 ($1{,}355$ women).
\textbf{Implementation.} Baselines use their official training code and
hyperparameters except batch size (per GPU, DDP on $8{\times}$A100 80GB: LMV-Net 8,
OA-BreaCR 24; ours 64 in Stage~1, Stage~2 on one GPU). All use a fully trainable ResNet-18 on $2048{\times}1664$ 8-bit grayscale inputs
(replicated to 3 channels), initialised from Mirai~\cite{yala2021mirai} (ours: unfrozen
again in Stage~2) or, for OA-BreaCR, ImageNet-1K as originally;
LMV-Net uses a frozen MammoRegNet~\cite{thrun2025reconsidering,thrun2025impact}
only for registration fields. Our Stage~2 warps priors with fields
precomputed by our dual-view registration method~\cite{krasnova2026jiim}.
\textbf{External reference.} Mirai~\cite{yala2021mirai} is evaluated with its
published weights (trained on about 211k MGH exams) and native 16-bit
preprocessing and normalisation, without fine-tuning. Its single per-exam risk
is assigned to both breasts, and metrics are computed per breast. A Karolinska
cohort that may overlap with ours served only as its external test set, so
the weights were not trained on our data and there is no leakage.
\textbf{Evaluation.} All metrics come from our common code, which we will release. The C-index is
the time-dependent IPCW C-index~\cite{uno2011cstat} as implemented by
OA-BreaCR~\cite{wang2024oabreacr}, on $F_\tau$ (Kaplan--Meier weights
$1/G(\tau)^2$, ties $\tfrac12$). 95\% CIs: 1,000 patient-level bootstrap resamples.

\section{Results}
\label{sec:results}

\textbf{High discrimination, no timing.} LMV-Net and OA-BreaCR reach C-indices
of 0.75 and 0.69 and AUC$_k$ of 0.66--0.77 (Table~\ref{tab:main}) even
without same-year diagnoses. Yet
their \ctime{} is 0.51 and 0.55, with CIs covering 0.5: neither orders cancer
cases by time to diagnosis significantly better than chance. The constant
reference scores better in \wmae{} and WCS (1.20 vs.\ 1.48--1.52, 60\% vs.\
42--46\%), as expected (Sec.~\ref{ssec:metrics}).

\begin{figure}[t]
\centering
\includegraphics[width=0.85\linewidth]{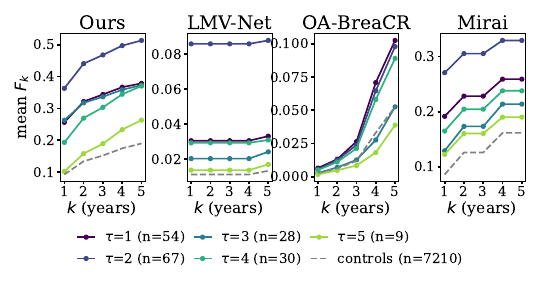}
\caption{Mean predicted cumulative risk $F_k$ on the test set, by true
year to cancer $\tau$ for cases, and for controls (dashed). Y-axes differ
between models.}
\label{fig:curves}
\end{figure}

\textbf{Risk-curve diagnostics.} For LMV-Net, the correlation of $F_1$ and
$F_5$ across exams is 0.999: the five outputs carry a single ranking, which is
the regime of the Proposition. Its five AUC$_k$ thus score one ranking against
five outcome definitions. OA-BreaCR, whose head is not subject to the
collapse, gives non-degenerate curves, but its mean $\that$ stays near 4.1
years for every true year (Fig.~\ref{fig:curves}). For OA-BreaCR, 96\% of
the case probability lies after year~1, and its most likely year is correct for only 7\% of cases (20\% for a uniform
guess). Regressing $\that$ on $\tau$ over cases (bootstrap over women) gives
a slope of 0.15 years per year (95\% CI 0.06--0.24) for our model, 0.07
($-0.05$--0.18) for LMV-Net, 0.015 ($-0.004$--0.032) for OA-BreaCR and
0.09 (0.02--0.16) for Mirai.

\textbf{Proposed method.} Our model reaches a C-index of 0.745 and AUC$_k$ of
0.72--0.77, on par with LMV-Net (overlapping CIs; both breast-level). It is the only model trained on CSAW-CC whose \ctime{} CI excludes 0.5 (0.58,
0.52--0.64), and its \wmae{} (1.39) is the lowest among trained models. It
does not reach the constant reference (1.20, 60\%), and its WCS (44.0\%) is
below LMV-Net's (45.9\%). Stage~1 alone has a constant year shape, so its \ctime{} is 0.5 by
construction; Stage~2 yields above-chance timing (95\% CI 0.52--0.64) and slightly raises the C-index (0.734 $\to$ 0.745).
A paired patient-level bootstrap (1,000 resamples, same draws for all
models) of the \ctime{} difference gives $+0.071$ vs.\ LMV-Net (95\% CI
0.008--0.130, one-sided $p=0.014$), $+0.034$ vs.\ OA-BreaCR
($-0.046$--0.107, $p=0.20$) and $+0.010$ vs.\ Mirai ($-0.061$--0.074,
$p=0.40$). Only the gain over LMV-Net is significant.
Only 72 of the 13{,}087 case pairs come from the same woman. On pairs from
different women, our \ctime{} is 0.580, so the score is not driven by
within-woman pairs.

\textbf{External reference.} Mirai matches our model in C-index (0.745) and
\ctime{} (0.57 vs.\ 0.58, paired difference not significant), despite its
partly collapsed head (Sec.~\ref{ssec:collapse}).

\section{Discussion and Conclusion}
\label{sec:conclusion}

Our results concern the shape of the predicted risk curves, not the ranking
of women by risk, which AUC$_k$ and C-index capture. They barely register
whether a model knows which cancers come first. Our model and LMV-Net
illustrate this: their C-indices are nearly identical (0.745 and 0.746), yet our \ctime{}
is higher (0.58 vs.\ 0.51, paired CI 0.01--0.13). Both start from the Mirai
encoder and are fine-tuned on CSAW-CC. LMV-Net, with an additive head, ends
with near-chance timing, whereas our two-stage model reaches the level of
Mirai's published weights (0.57). On
CSAW-CC, neither of two recent longitudinal models orders cancer cases
significantly better than chance. We recommend reporting \ctime{} (and year errors only next
to a constant reference) whenever per-year risks are claimed, and checking
that per-year ReLU increments stay active.

\textbf{Limitations.} We use a single dataset (CSAW-CC). The diagnosis year
$\tau$ has annual granularity and, as in~\cite{thrun2026lmvnet}, is derived
from the CSAW-CC radiological timing field, so it may contain label noise. We
assess ranking and curve shape, not the calibration of $F_k$. No trained model
reaches the constant reference in year error. Future work includes training
on EMBED~\cite{jeong2023embed} and evaluation across datasets.

\section*{Compliance with ethical standards}
\label{sec:ethics}
This research study was conducted retrospectively using human subject data from the CSAW-CC dataset, made available for research by Karolinska Institutet. The collection and use of the data for AI research were approved by the Swedish Ethical Review Authority (Dnr 2019-01946, 2021-01030, 2019-03638) and the Regional Ethical Review Board in Stockholm (Dnr 2016/\allowbreak 2600-31), which waived the re\-quire\-ment for informed consent.

\section*{Acknowledgments}
\label{sec:acknowledgments}
The research was carried out using the MSU-270 super\-computer of Lomonosov
Moscow State University. We thank the authors of
OA-BreaCR~\cite{wang2024oabreacr} and
LMV-Net~\cite{thrun2026lmvnet} for releasing their code, and the authors of
Mirai~\cite{yala2021mirai} for releasing their code and model weights.
The study was conducted under the state assignment of Lomonosov Moscow State
University. The authors have no relevant financial or
non-financial interests to disclose.
AI-based tools were used for language editing and coding assistance. The authors reviewed all outputs and take full responsibility for the content.

\bibliographystyle{IEEEbib}
\bibliography{references}

\end{document}